\documentclass[12pt]{article}
\usepackage{psfig}

\begin{document}

\title{\bf Charm particle production in hadronic collisions}

\author{{C. Avila$^{1)}$\footnote{cavila@uniandes.edu.co}, 
J. Magnin$^{1,2)}$\footnote{jmagnin@cbpf.br}, 
L.M. Mendoza-Navas$^{1)}$}\footnote{luismi@fnal.gov}\\
\small{$^{1)}$ {\it Departamento de F\'{\i}sica, Universidad de los Andes}}\\
\small{A.A. 4976, Santaf\'e de Bogot\'a, Colombia}\\
\small{$^{2)}$ {\it Centro Brasileiro de Pesquisas F\'{\i}sicas}}\\
\small{Rua Dr. Xavier Sigaud 150, Urca, CEP 22290-180, Rio de Janeiro, Brazil}}

\maketitle

\begin{abstract}
We study charm particle production in hadron-hadron collisions. 
After calculating perturbatively the charm quark differential cross section, we 
study the hadronization mechanisms. It is shown that recombination is the key to 
understand the so called Leading Particle Effect in charm meson production. For 
charm baryon production, however, leading particle effects are due to a combination 
of contributions coming from both, the recombination and the fragmentation mechanisms. 
We compare our calculations to experimental data on charm hadron production in 
$\pi^-~N$ and $p~N$ interactions from several experiments 
and show that a consistent description of them can be reached without the aid of 
other mechanism than recombination and fragmentation.
\end{abstract}

\newpage

\section{Introduction}

Heavy quark production in hadronic collisions is one of the most interesting testing grounds 
of Quantum Chromodinamics. The fusion reactions $g~g \rightarrow Q~\bar Q$ and 
$q~\bar q \rightarrow Q~\bar Q$ are expected to be dominant in heavy quark, $Q$,  
production and, as a matter of fact experimental data seems to
reasonably agree with perturbative QCD (pQCD) calculations which today
are available up to Next to Leading Order (NLO)~\cite{nason}. 

Once a heavy quark is produced, it has to hadronize to produce the
observed, hadronic, final state. 
On this respect, it was expected that the factorization theorem be valid and 
hadronization proceeded 
through the fragmentation mechanism. Thus, heavy hadron production can be separated into the 
perturbatively calculable hard scattering and gluon dynamics from the non-perturbative 
bound state dynamics contained in the process independent hadron structure, expresed through 
the corresponding parton distribution functions (PDF), $q(x,Q^2)$ and $g(x,Q^2)$, and the jet 
fragmentation functions, $D_{h/Q}(z,Q^2)$~\footnote{$x$ is the
  momentum fraction of partons inside the initial hadron, $z$ is the
  momentum fraction of the heavy quark in the final hadron. $Q^2$ is
  the momentum transfer.}. 
Literally speaking, the factorization assumption 
predicts strict independence of the heavy quark hadronization from the production 
process. Thus, no flavor correlation should exist between initial and final states.

However, there exist copiuos experimental information on charm hadron production 
which contradicts the above
hypothesis~\cite{e791-baryon,selex,wa92,wa82,alves1,e791-2,
alves,e791}. In fact, there was observed an excess in the 
charm hadron production at large values of $x_F$ ($\sim2p_l/\sqrt{s}$) and a correlation 
between the leading\footnote{A leading particle is defined as the one which shares valence quarks
with the colliding hadrons} charm hadrons with the projectile quantum numbers. This suggests the 
presence of other hadronization mechanisms which must be relevant at large values of 
$x_F$ and low $p_T^2$, as long as fragmentation of the charm quark should not produce either 
flavor correlations between initial and final states, neither charm
hadrons at large values of $x_F$. 
In particular, flavor correlation suggests hadronization mechanims by which projetile 
spectators produced at small $p_T^2$ recombine with charm quarks produced either perturbatively in 
the hard QCD process, or charm quarks, though perhaps of a non-perturbative nature, already 
present in the structure of the beam particles.

From a theoretical point of view, models have been proposed to account for 
the enhancement of charm hadron production at large $x_F$ and flavor correlations. 
Among them we can mention the model of Ref~\cite{kartvelishvili}, in which a charm quark produced
perturbatively recombines with the debris of the projectile, the intrinsic charm 
model~\cite{brodsky}, in which a particular Fock state of the projectile containing 
charm quarks breaks in the collision giving thus the desired flavor correlation between 
initial and final particles, recombination type models~\cite{magnin} in which charm quarks 
already present in the projectile structure recombine with light quarks, and models based 
in the Dual Parton Model and Dual Topological Unitarization~\cite{piskunova} in which 
both, the heavy quark production and hadronization, are treated on a non-perturbative basis. 
Likewise, most recently two new approachs based in recombination have been presented. 
In the first, recombination of charm and light quarks, both being part of a hard QCD diagram, 
recombine to produce a charm hadron~\cite{braten}. In this scheme, recombination is thought as 
a higher order correction to hadronization since hard scattering diagrams contributing to 
this process are of NLO or higher. The second approach~\cite{rapp} involves a modification of 
the usual recombination prescription to produce an enhancement in the central (low $x_F$) 
production of charm hadrons. 

All the above models have been more or less successful in reproducing the main features of 
charm hadron production, with possibly the only exception of the intrinsic charm 
model~\cite{brodsky}, which seems to be excluded~\cite{anjos} by recent experimental 
data on charm baryon production in $\pi^-N$ interactions by the E791~\cite{e791-baryon} 
and SELEX~\cite{selex} Collaborations. However, it is important to remark that, although some 
models are able to reproduce experimental data on production asymmetries, they cannot reproduce 
simultaneously data on production asymmetries and differential cross sections. This is the case 
of the intrinsic charm model, as noted in Ref.~\cite{anjos}. This shows that in order to 
make a meaningful comparison among models and experimental data, both, the differential cross 
section and the production asymmetry have to be taken properly into
account. As a matter of fact, we 
have two out of three quantities which are independent, namely, the differential 
cross sections of both, particle and antiparticle, or one of the cross sections and the 
production asymmetry.

In what follows we shall analyse the main features of perturbative charm production in 
hadron-hadron collisions followed by the study of possible hadronization mechanisms. 
Later, we will compare model results with available experimental data. The last section 
will be devoted to conclusions.

\section{Brief review of perturbative charm production}

In the parton model, charm quarks are produced through the interaction of 
partons in the initial hadrons. The differential cross section as a function of $x_F$ 
is given by~\cite{vbh-npb}
\begin{equation} 
\frac{d\sigma_{c(\bar c)}}{dx_F}=\frac{1}{2} \sqrt{s} 
\int H_{ab}(x_a,x_b,\mu_F^2,\mu_R^2) \frac{1}{E} dp_T^2 dy \: ,
\label{1.1} 
\end{equation} 
where $H_{ab}$ is a function containing information on the structure of the colliding 
hadrons $a,b$, and on the hard QCD process which produces the charm quarks. At LO, the funtion 
$H_{ab}$ reads
\begin{eqnarray}
H_{ab}(x_a,x_b,\mu_F^2,\mu_R^2)& = & \Sigma_{i} \left( q_i~^a(x_a,\mu_F^2)
\bar{q_i}~^b(x_b,\mu_F^2) \right. \nonumber \\
                   &   & + \left. \bar{q_i}~^a(x_a,\mu_F^2) q_i~^b(x_b,\mu_F^2) 
\right) \frac{d \hat{\sigma}}{d
\hat{t}} \mid_{q\bar{q}}(\hat{s},m_c,\mu_R^2) \nonumber \\
                   &   & + g_i~^a(x_a,\mu_F^2) g_i~^b(x_b,\mu_F^2) 
\frac{d \hat{\sigma}}{d \hat{t}}\mid_{gg}(\hat{s},m_c,\mu_R^2),
\label{1.2}
\end{eqnarray}
with $x_a$ and $x_b$ being the parton momentum fractions, $q(x,\mu_F^2)$ and
$g(x,\mu_F^2)$ the quark and gluon momentum distributions in the colliding particles, 
$\hat{s}=x_ax_bs$ is the square of c.m. energy of the partonic system
and $\mu_F$ and $\mu_R$ are the factorization and the renormalization
scales respectively. In eq.~(\ref{1.1}), $p_T ^2$ is the 
squared transverse momentum of the produced $c$-quark, $y$ is the rapidity 
of the $\bar {c}$ quark and $E$ the energy of the produced $c$-quark. The sum 
in eq.~(\ref{1.2}) runs over 
$i = u,\bar{u},d,\bar{d},s,\bar{s}$.

\begin{figure}[t]
\centerline{\psfig{figure=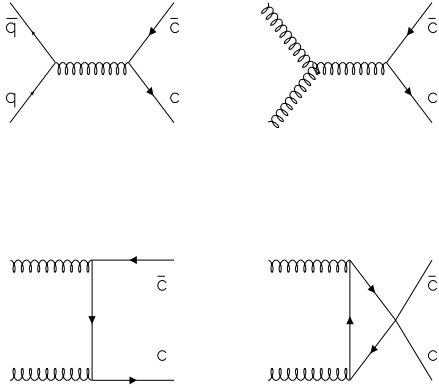,height=3.0in}}
\caption{Feynman diagrams for the elementary cross sections entering in eqs.~(\ref{1.3}) and 
(\ref{1.4}) at LO.}
\label{fig1}
\end{figure}

The elementary cross-sections for $q-\bar{q}$ anihilation and gluon
fusion, $d\hat{\sigma}/d \hat{t}\mid_{q \bar {q}}$ and $d \hat{\sigma}/d
\hat{t}\mid_{gg}$ respectively, at LO are given by~\cite{vbh-npb,combridge}.
\begin{equation} \frac{d
\hat{\sigma}}{d \hat{t}}\mid_{q \bar {q}} = \frac{\pi \alpha _{s}^{2}
 \left( \mu_R^2 \right)}{9 \hat{m}_{c}^{4}} \;  \frac{cosh \left( \Delta y
\right) + m_{c}^{2}/ \hat{m}_{c}^{2}} {\left[ 1+cosh \left( \Delta y
\right) \right] ^3} 
\label{1.3} 
\end{equation} 
\begin{equation}
\frac{d \hat{\sigma}}{d \hat{t}}\mid_{gg}= \frac{\pi \alpha_{s}^{2} \left(
\mu_R^2 \right)}{96 \hat{m}_{c}^{4}} \;  \frac{8 cosh \left( \Delta y \right)
-1}{\left[ 1+cosh \left( \Delta y \right) \right]^3} \: \left[ cosh \left(
\Delta y \right)+
\frac{2m_c^2}{\hat{m}_c^2}+\frac{2m_c^4}{\hat{m}_c^4}\right], 
\label{1.4}
\end{equation} 
where $\Delta y$ is the rapidity gap between the produced
$c$ and $\bar{c}$ quarks and $\hat{m}_c^2=m_c^2+p_T^2$.  The Feynman
diagrams involved in the calculation of eqs.~(\ref{1.3}) and (\ref{1.4})
are shown in Fig. \ref{fig1}.

As shown in eqs.~(\ref{1.3}-\ref{1.4}), at LO the only dependence on $\mu_R$  
in the elementary cross sections is in $\alpha_{s}^{2} \left(\mu_R^2 \right)$. 
At this order, the value of the renormalization scale is fixed by the requirement that  
the propagators in diagrams of Fig.~\ref{fig1} be off-shell by a quantity of at least 
$m_c^2$. So, it is common to use $\mu_R^2 = m_c^2$. 

Concerning the factorization scale $\mu_F$, it is the scale at which
the initial hadrons ``see'' one another in the collision. Then, in the 
above sense, $\mu_F$ gives the quark and gluon content of the initial 
hadrons at the time of the collision. However, differently to the
case of the renormalization scale $\mu_R$, there is no guiding
principle to help fixing $\mu_F$. Along this
work, and in order to avoid problems related with flavor excitation diagrams
containing heavy quarks~\cite{nason,combridge}, we shall use $\mu_F=1$
GeV, below the threshold for charm production and consistent with the
sum over light flavors in eq.~(\ref{1.2}).

It is also known that NLO and LO calculations only differ by a $K$-factor of the order 
of $2-3$~\cite{frixione}, meaning that calculations can be done
consistently at LO and then multiplied by the corresponding
$K$-factor. NLO calculations also show a tiny $c-\bar c$ asymmetry~\cite{nason}, 
which is too small to produce any effect after hadronization. Note however that, although the 
small $c-\bar c$ asymmetry arising at NLO has the same sign that the observed $D^+-D^-$ 
production asymmetry (i.e. $\bar c$ and $D^-$ favored over $c$ and $D^+$ production), it 
has the opposite sign for baryon production.

\section{Hadronization mechanisms}

\begin{figure}[t]
\begin{minipage}{2.5in}
\psfig{figure=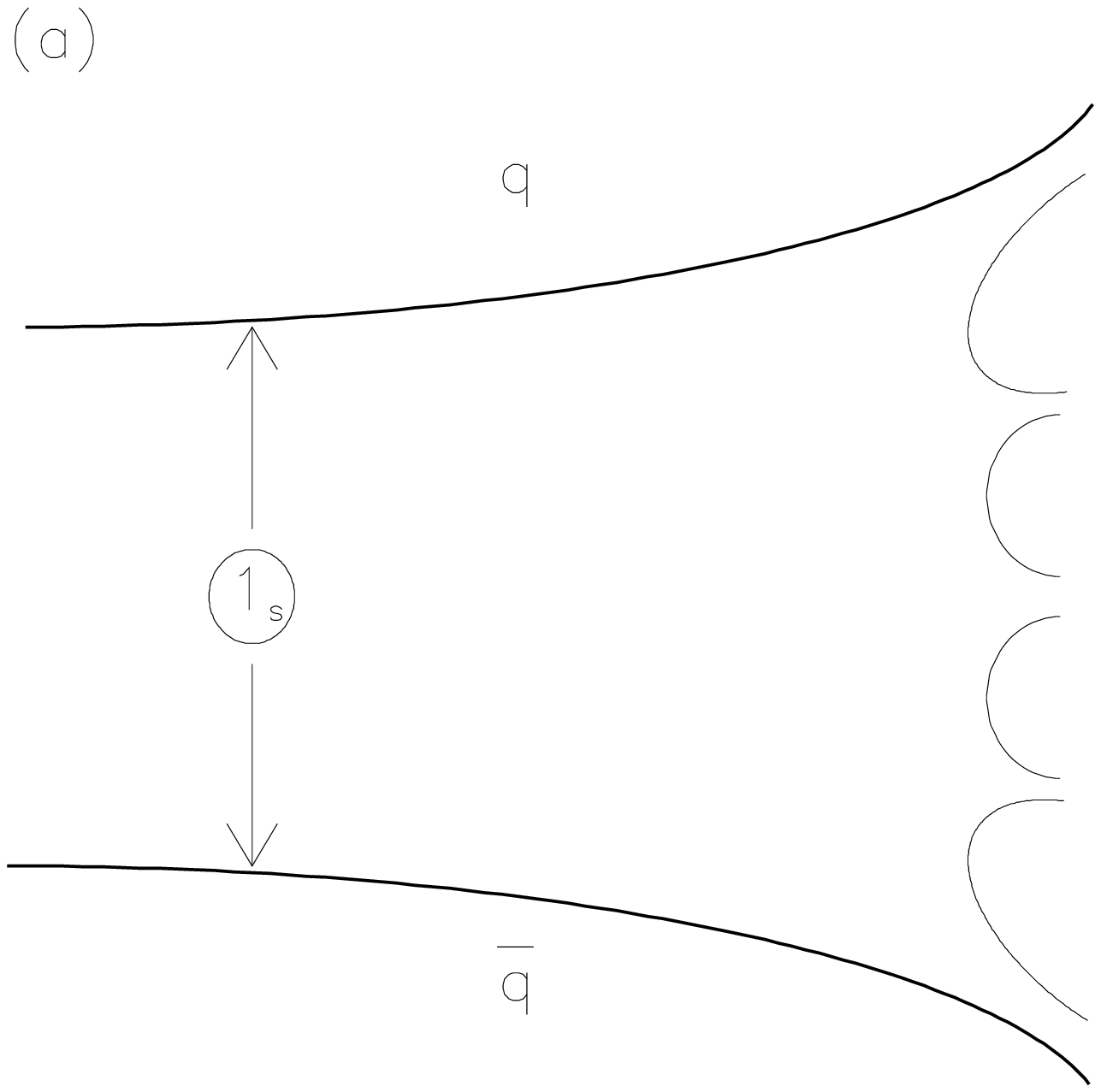,height=3.0in}
\end{minipage}
\begin{minipage}{2.5in}
\psfig{figure=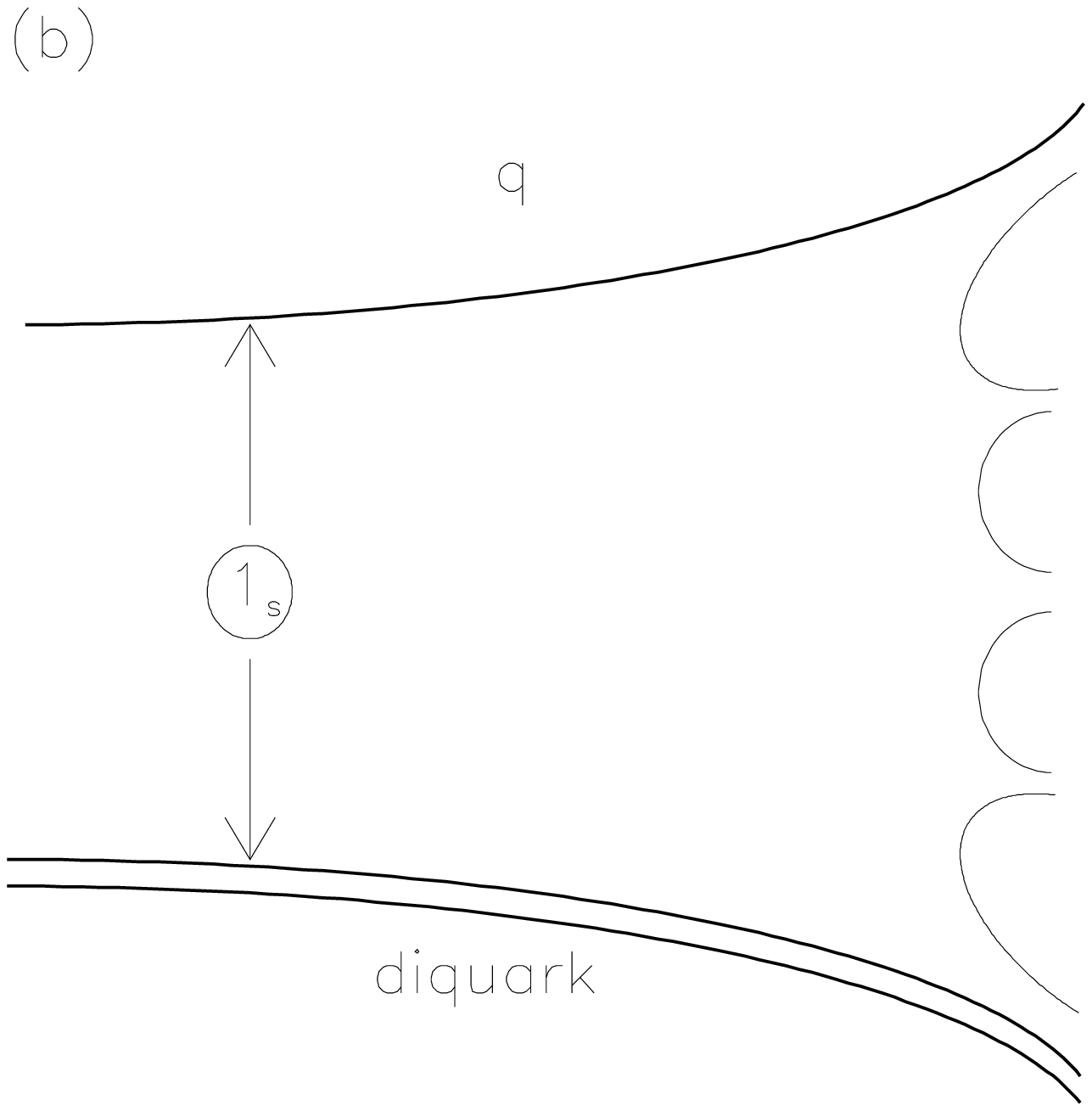,height=3.0in}
\end{minipage}
\caption{Allowed color string configurations: meson-like string (a) and baryon-like string (b).}
\label{fig3}
\end{figure}

There are basically two different models for charm quark hadronization, namely,
recombination and fragmentation. In the first, the $c~(\bar{c})$-quark
produced in a pQCD process joints to the debris of the initial
particles to form the final charm hadron~\cite{kartvelishvili} while in the second, the
$c~(\bar{c})$-quark fragments to the final charm hadron leaving a
string of quarks behind it~\cite{peterson}. Variation of these
processes can also be found in the
literature~\cite{variations}. However, there are some features which are
common to any class of hadronization processes: hadrons are colorless,
which means that hadronizing quarks must be in a color singlet state
and, at the end of hadronization, no free quarks could exist any
more. These two basic requirements are the consequence of confinement
in QCD. Consequently, whatever 
the hadronization process be, a color string must be formed among the
hadronizing quarks which has to have the correct
color quantum numbers and the exact number of quarks to ensure that
no free quarks remain at the end of the process and colorless hadrons
are formed. These lead us to the following clasification: {\it i)
  Short color strings}, responsible for recombination processes, in
which a $q-\bar q$ pair or a $qqq$ forms the final hadron without
further hadron emission, and {\it ii) Large color strings},
producing the final hadron by fragmentation being it accompanied by
the emission of light (mostly pion) mesons. Note that, in addition,
for each case there exist also two different types of color strings,
namely, the meson and the baryon-like strings, being them
characterized by the fact that the first is formed among a quark and
an antiquark and the second one between a quark and a diquark, as
shown in Fig.~\ref{fig3}. The first
will produce predominantly mesons while the later has to produce at
least one baryon in order to conserve the baryon number. 

In what follows, we will analyze both, fragmentation and recombination,
having in mind their applicability to heavy quark hadronization process.

\subsection{Fragmentation}

\begin{figure}[t]
\centerline{\psfig{figure=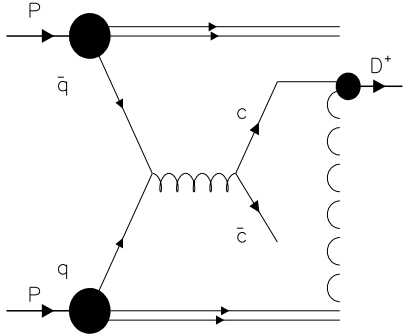,height=3.0in}}
\caption{Typical configuration leading to $D^+$ production in $p-p$ interactions by 
color string fragmentation.}
\label{fig4}
\end{figure}

A typical quark configuration leading to fragmentation is shown in 
Fig.~\ref{fig4}.
Formation of color strings among the charm quarks and the remnants of 
the colliding hadrons would give rise to the production of open charm, i.e. $D$ 
mesons and eventually charm baryons. 

This kind of contribution to hadronization is usually modeled by the convolution 
of the $c$-quark differential cross section with a Peterson Fragmentation 
Function, $D_{H/c}$, \cite{peterson}, 
\begin{eqnarray}
\frac{d\sigma_H}{dx_F} &=& \int{\frac{dz}{z}\frac{d\sigma_{c(\bar c)}}
{dx}D_{H/c}(z)}\; , \nonumber \\ 
z &=& \frac{x_F}{x}\; , \nonumber \\
D_{H/c}(z) &=& \frac{N}{z\left( 1 - \frac{1}{z} - \frac{\epsilon}{1-z} 
\right)^2} \; ,
\label{3.1}
\end{eqnarray}
where $d\sigma_{c(\bar c)}/dx$ is given in eq.~(\ref{1.1}). 
Along this work we will use $\epsilon=0.022$ for $D$ meson production and 
$\epsilon=0.06$ for $\Lambda_c$. These values for $\epsilon$ have been extracted from 
charm hadron production in $e^+-e^-$ interactions~\cite{pdg}. Note
that the fragmentation function is used to model the production of a
hadron containing a heavy quark at the end of the color string,
regardless of what happens with the remaining of the color string.

\subsection{Recombination}

\begin{figure}[t]
\centerline{\psfig{figure=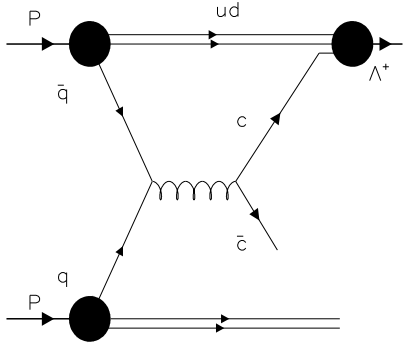,height=3.0in}}
\caption{Typical configuration leading to $\Lambda_c^+$ production in $p-p$ interactions by 
recombination.}
\label{fig5}
\end{figure}

A typical quark configuration leading to recombination is shown in Fig.~\ref{fig5}.
The contribution of the recombination mechanism can be estimated by means 
of~\cite{kartvelishvili}
\begin{equation}
\frac{d\sigma}{dx_F} = \frac{\sqrt{s}}{2}\int{x_vz^\star\frac{d\sigma^\star}
{dx_vdz}~R(x_v,z,x_F)\frac{dx_v}{x_v}\frac{dz}{z}}\; ,
\label{3.2}
\end{equation}
where $R(x_v,z,x_F)$ is the recombination function, for which we shall use~\cite{das-hwa}
\begin{equation}
R\left( x_v,z,x_F\right) = \beta\frac{x_v~z}{x_F^2}\delta\left(1-\frac{x_v+z}{x_F}
\right)\; ,
\label{3.3}
\end{equation}
with $\beta$ a normalization parameter which has to be fixed from 
experimental data. In eq.~(\ref{3.2}), $z^\star=2E_c/\sqrt{s}$ and  
\begin{eqnarray}
\frac{d\sigma^\star}{dx_vdz} &=& q_v(x_v,\mu_F^2)\frac{d\hat{\sigma}}{dx_vdz}\; ,\nonumber \\
\frac{d\hat{\sigma}}{dx_vdz} &=& 
\int_{0}^W{dp_T^2 \int_{z_+/(1-z_-)}^{1-x_v}{\frac{H_{ab}(x_a,x_b,\mu_F^2,\mu_R^2)}
{x_a-z_+}}~dx_a}\; ,
\label{3.4}
\end{eqnarray}
where $H_{ab}$ is given in eq.~(\ref{1.2}), $x_v$ is the fraction of the momentum of the 
hadron $a$ carried by the spectator quark $q_v$, 
$z=2p_{z,c}/\sqrt{s}$, $z_\pm = \frac{1}{2}(z^\star \pm z)$, 
$x_b=x_az_-/(x_a-z_+)$ and 
\begin{eqnarray}
E_c & = & \sqrt{m_T^2+p_{z,c}^2}\; ,\nonumber \\
W & = & \frac{s(1-x_v-z)(1-x_v)(1+z)}{(2-x_v)^2}-m_c^2\; .
\label{3.4a}
\end{eqnarray}

In eq.~(\ref{3.4}), $q_v$ is the probability density function of a
quark in the initial hadron. For baryon production, $q_v$ has to be replaced by 
$q_v(x_{v1},x_{v2},\mu^2_F) =
q_{v1}(x_{v1},\mu^2_F)~q_{v2}(x_{v2},\mu_F^2)$ and the recombination
function of eq.~(\ref{3.3}) by~\cite{ranft}
\begin{equation}
R\left( x,y,z,x_F\right) = \beta\frac{x~y~z}{x_F^3}\delta\left(1-\frac{x+y+z}{x_F}
\right)\; .
\label{3.3b}
\end{equation}
Notice that, as the quark $q_v$ contains information about the structure
of one of the colliding hadrons, this introduces a flavor correlation
among the initial and final states.

\begin{figure}[t]
\centerline{\psfig{figure=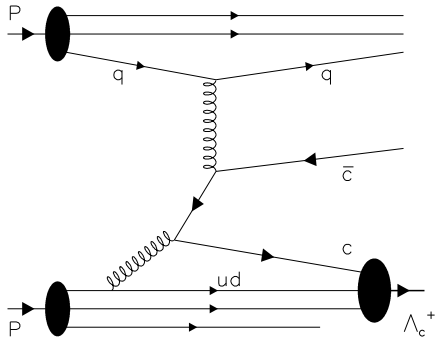,height=3.0in}}
\caption{Typical configuration leading to $\Lambda_c^+$ production in $p-p$ interactions by 
spectator recombination.}
\label{fig6}
\end{figure}

\subsection{Direct recombination}
\label{dirrec}

A second  possibility is that a {\it spectator} charm quark in a
flavor excitation subdiagram recombines with the debris of one of the
initial hadrons as shown in  Fig.~\ref{fig6}. This kind of
contributions can be calculated along the lines of the model of
Refs.~\cite{magnin,luismi} in which a charm quark
being part of the sea of the initial hadron recombines with valence
and sea quarks after the collision. Notice, however, that the
factorization scale must be consistent with the fact
that charm quarks exist in the sea of the initial hadron, i.e. $\mu_F^2
\ge m_c^2$. The production of charm mesons is then given by~\cite{das-hwa}

\begin{equation}  
\frac{d\sigma^{rec}}{dx_F}=\beta
\int_0^{x_F}\frac{dx_1}{x_1}\frac{dx_2}{x_2}F_2\left( x_1,x_2\right) 
R_2\left( x_1,x_2,x_F\right),
\label{3.5} 
\end{equation}
where $x_i$, $i=1,2$, is the momentum
fraction of the $i^{th}$ quark, $F_2 \left( x_1,x_2 \right) $ is the
two-quark distribution function in the initial hadron and $R_2\left(x_1,x_2,x_F\right)$ 
is the two-quark recombination function with $\beta$ a normalization constant which must be 
fixed by comparison to experimental data. Baryon production can be
calculated just by replacing the $F_2$ and $R_2$ by the corresponding
functions for three quark recombination. Along this work we will use
\begin{eqnarray}
F_i(x_1,...,x_i) &=& \left[\Pi_{j=1}^i~f_j(x_j,\mu_F^2)\right]
(1-\sum_{j=1}^i x_j)^\gamma \nonumber \\
i &=&2,3\; ,
\label{3.6}
\end{eqnarray}
with $\gamma = 1$ for $i=2$ and $\gamma=-0.1$ for $i=3$~\cite{magnin} and,
for the recombination function we shall use its simplest version given by~\cite{das-hwa,ranft}
\begin{eqnarray}
R_i(x_1,...,x_i,x_f) &=& \frac{\Pi_{j=1}^ix_j}{x_f^{i}}\delta
\left(1-\frac{\sum_{j=1}^i x_j}{x_F}\right)\; . \nonumber \\
i &=&2,3
\label{3.7}
\end{eqnarray}

\section{Modeling and comparison to experimental data}

Several experiments have measured the differential cross
section both as a function of $x_F$ and $p_T^2$ for charm meson and also
baryon production. A few of them have also measured the production
asymmetry, defined by
\begin{equation}
A\left(x_F\right) = \frac{dN^L/dx_F - dN^{NL}/dx_F}{dN^L/dx_F + dN^{NL}/dx_F}\; ,
\label{comp2}
\end{equation}
where $L$ is for Leading and $NL$ is for Non Leading particles.

Along this section we shall compare the production model presented in
the preceding section with experimental data for $D$ meson production
in $\pi^--Nucleon$ interactions and for  $\Lambda_c$ production in 
$\pi$, $p-Nucleon$ interactions, which are, up to our knowledge, most
of the existing data in charm particle production. 

In order to compare the theoretical model to experimental data, we will
use 
\begin{equation}
\frac{d\sigma}{dx_F} = a \frac{d\sigma}{dx_F}^{frag} + b \frac{d\sigma}{dx_F}^{rec}\; ,
\label{comp1}
\end{equation}
where the first term in the right hand side of eq.~(\ref{comp1})
accounts for charm hadron production through fragmentation and its
expression is given in eqs.~(\ref{3.1}), and the second term represents 
the charm hadron production by the recombination of the perturbatively 
generated charm quark with spectator quarks, remnants of the beam 
particles, as given in eq.~(\ref{3.2}). 

We do not include any contribution coming from the direct recombination
of the debris of the initial particles (see Sec.~\ref{dirrec})
because it is naturally included when NLO contributions
to heavy quark production are taken properly into account, either by
including explicitly NLO diagrams in the perturbative part of the
process, either through the effective K-factor. However, we recognize
that as charm quarks are in the limit of what is understood as a heavy
quark, this mechanism is equally good to describe the recombination
part of the hadronization process (See e.g. Refs~\cite{magnin,anjos,luismi}).

Coefficients  $a$ and $b$ in eq.(\ref{comp1}) where fixed by fitting
experimental data. In most of the data, the differential cross section
for particle and antiparticle production were fitted independently 
by minimizing $\chi^2$ using the model of eq.~(\ref{comp1}). In these cases we used
\begin{equation}
\chi^2 = \sum_i{\left(\frac{y_i-\bar{y_i}}{\sigma_i}\right)^2}
\label{comp3}
\end{equation}
to obtain a linear system of equations in the unknowns $a$ and $b$ by
requiring that $\chi^2$ has a minimum and then, the resulting 2 $\times$ 2 linear
system of equations  was solved analitically. When fits involved one of the differential cross
sections and the asymmetry, the above procedure is not longer possible
since a non linear $4\times 4$ system of equations is obtained. In these cases we defined a
combined $\chi^2$ as the sum of the $\chi^2$ for the differential cross
section plus the one corresponding to the asymmetry and used MINUIT to
perform the fits.

We have to remark that particle and antiparticle $x_F$
distributions and the production asymmetry form a set of three
non-independent measurements. Then we performed fits to two of these 
measurements and derived the third from the other two.

In the following we analyse charm meson and baryon data and compare with our model in 
eq.~(\ref{comp1}).

\subsection{$D^\pm$ production}

\begin{figure}[t]
\begin{minipage}{2.0cm}
\psfig{figure=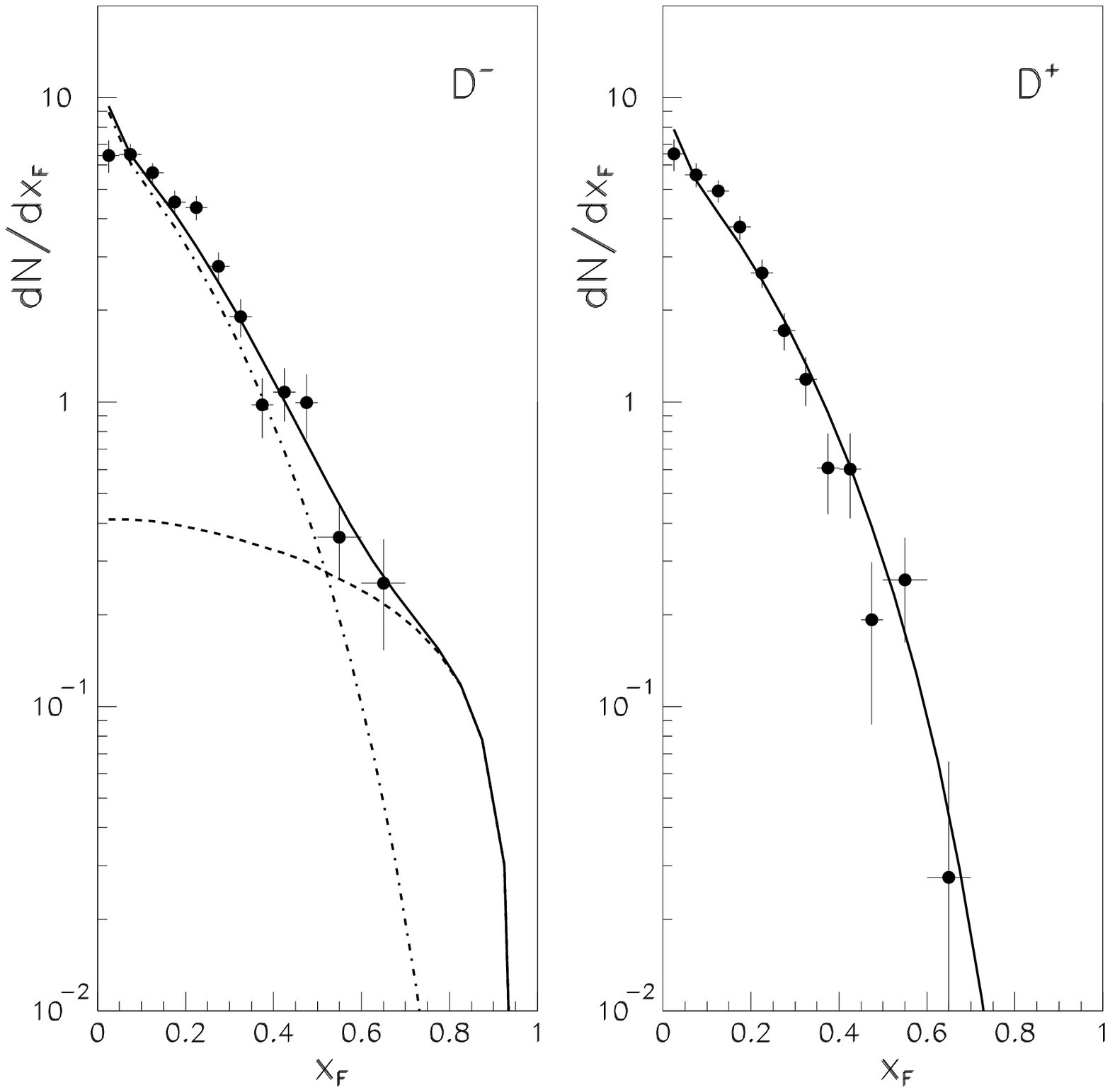,height=3.1in}
\end{minipage}
\begin{minipage}{7.5cm}
\psfig{figure=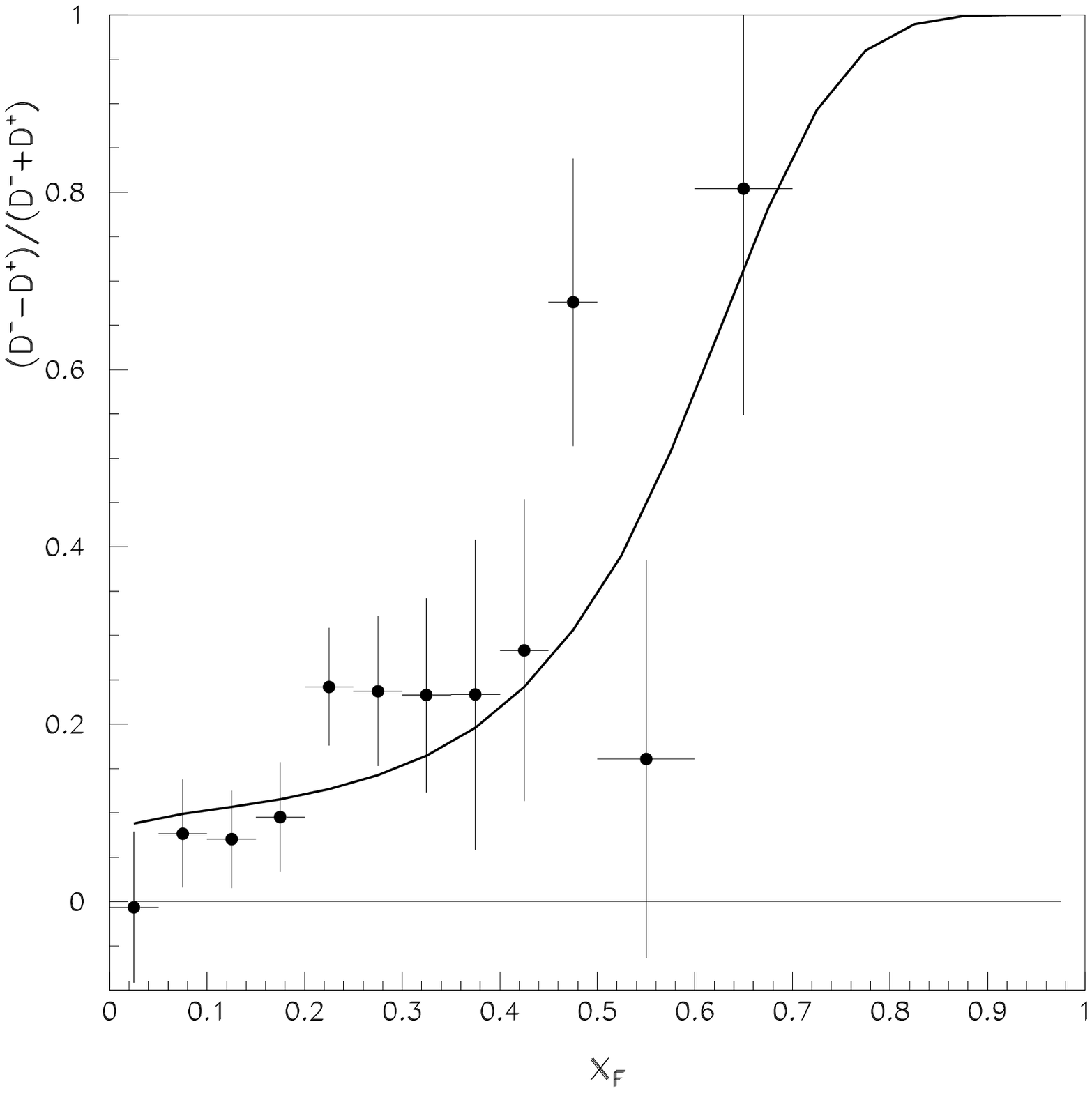,height=3.1in}
\end{minipage}
\caption{$D^-$ and $D^+$ production by the WA92 Collaboration~\cite{wa92}. 
Left and middle panels: (full line) our model as in eq.~(\ref{comp1}) compared to experimental 
data. Dashed and point-dashed lines: contributions from recombination
and fragmentation respectively. Right: Production
asymmetry. Model (full line) {\it vs} experimental data.}
\label{fig96}
\end{figure}

Charged $D$ meson production has been measured in $\pi^--Nucleus$ interactions at 350 
GeV/$c$ by the WA92 Experiment~\cite{wa92}, 340 GeV/$c$ by the WA82 Experiment~\cite{wa82}, 
 250 GeV/$c$ by the E769 Collaboration~\cite{alves1} and 500 GeV/$c$ by the 
E791 Collaboration~\cite{e791-2}. E769~\cite{alves} and E791~\cite{e791} Collaborations 
also measured $D^{*\pm}$ production in 250 and 500 GeV/$c$ $\pi^--Nucleus$ interactions 
respectively. 

In Fig.~\ref{fig96}  we display the results of our fit to experimental 
data on $D^\pm$ production by the WA92 Collaboration~\cite{wa92}. The WA92 experiment used 
targets of copper and tungsten. As can be seen in the figures, the model describes quite well the 
experimental data. Our fit also shows that charm quark fragmentation and recombination of the 
charm quark produced in the hard QCD process with the debris of the initial beam particles is 
enough to describe the data. The asymmetry, as shown by the curve in
Fig.~\ref{fig96}, has been calculated from the curves 
obtained from fits to particle and antiparticle distributions using eq.~(\ref{comp2}).

\begin{figure}[t]
\begin{minipage}{2.0cm}
\psfig{figure=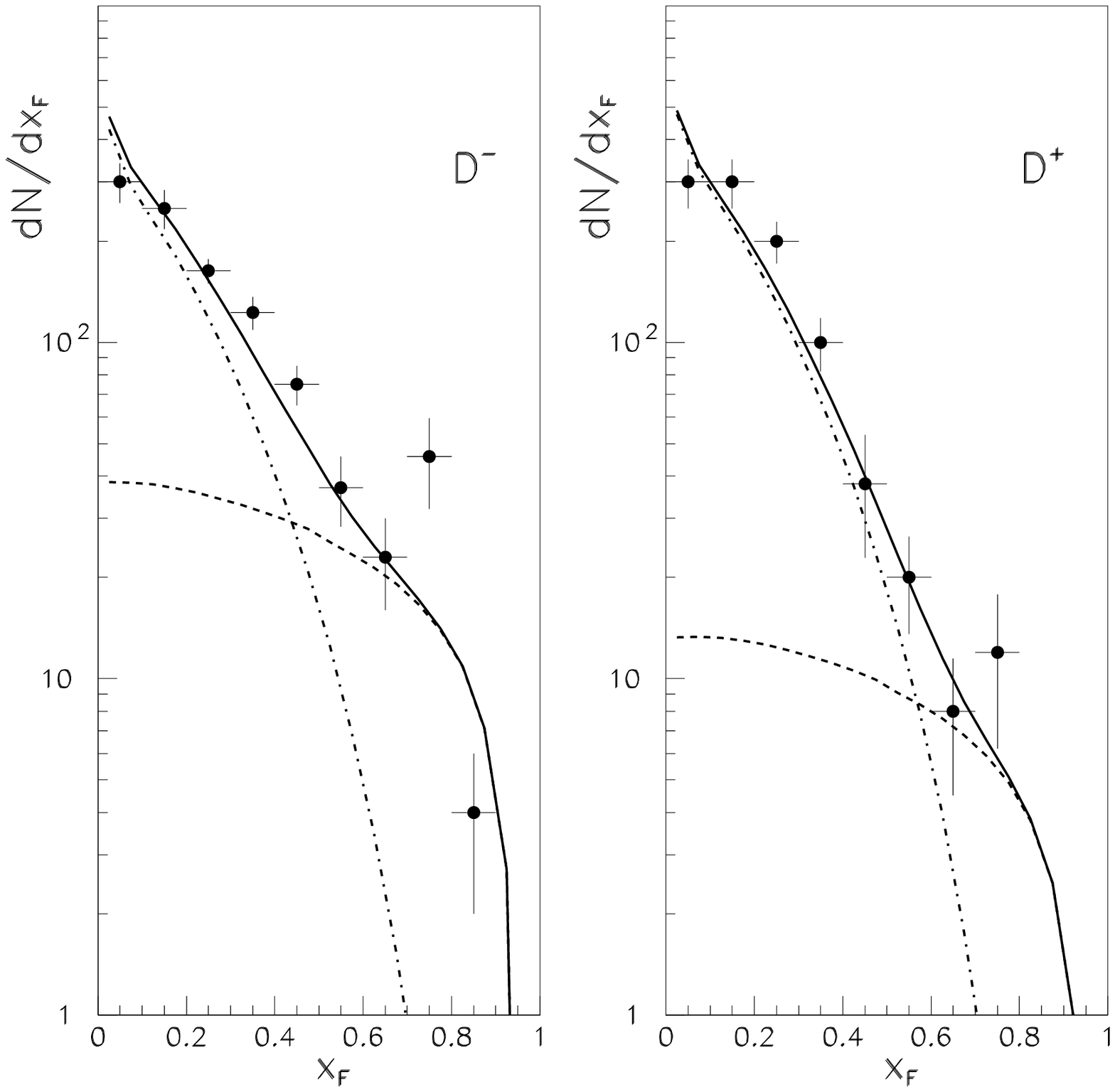,height=3.1in}
\end{minipage}
\begin{minipage}{7.5cm}
\psfig{figure=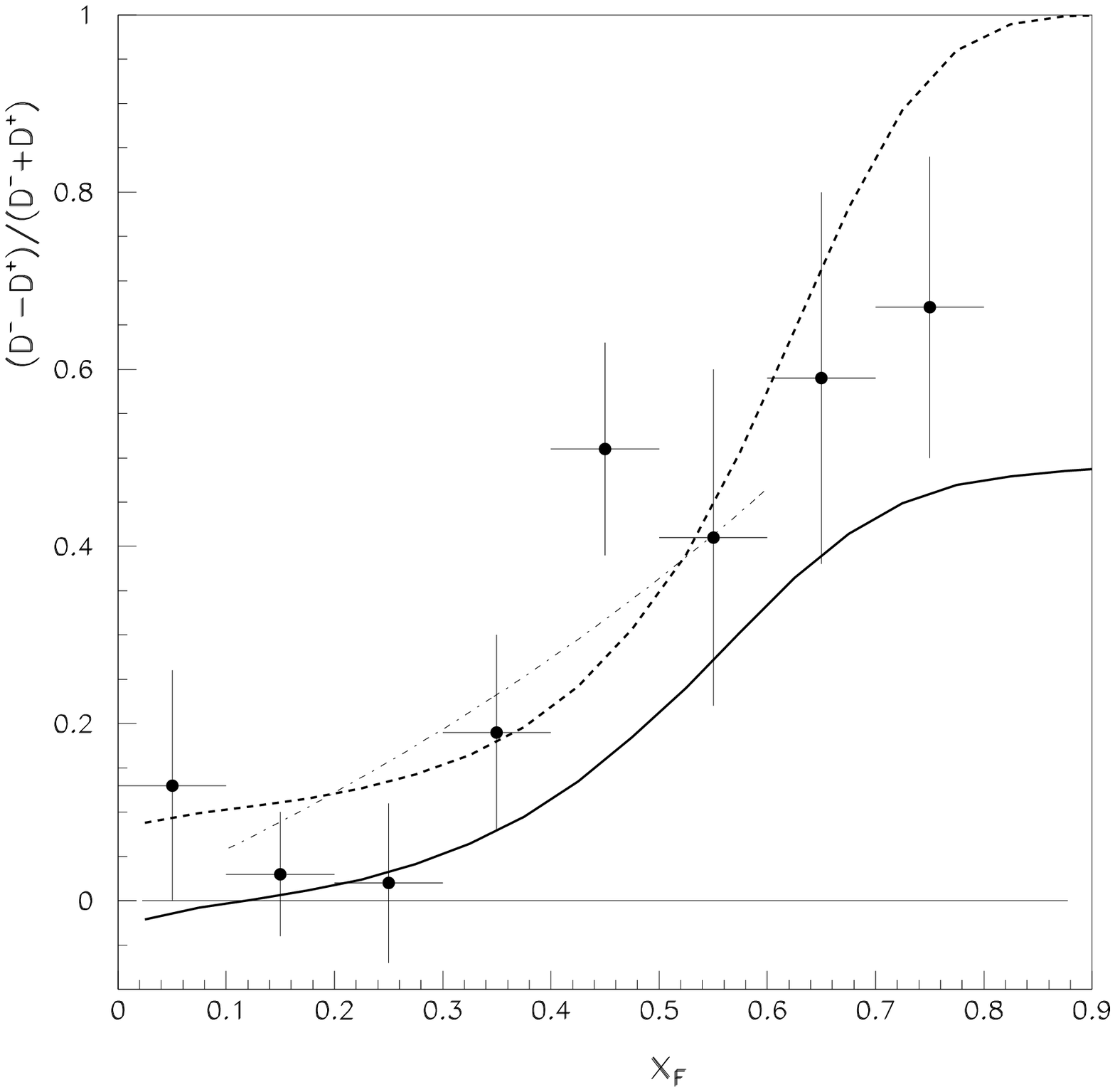,height=3.1in}
\end{minipage}
\caption{Left: $D^-$ and $D^+$ production by the WA82 Collaboration~\cite{wa82}. 
Left and middle panels: (full line) our model as in eq.~(\ref{comp1}) compared to experimental 
data. Dashed and point-dashed lines: contributions from recombination
and fragmentation respectively. Right: Production asymmetry. Model
(full line) {\it vs} experimental data. 
Also shown te asymmetry as fitted to the WA92 data (dashed line) and
the one extracted from measurements of the $D^-$ and $D^+$ $x_F$
distributions by the E769 Collaboration~\cite{alves1} (dashed-point line).}
\label{fig97}
\end{figure}

In Fig.~\ref{fig97} we display the results of our fit to experimental data on $D^\pm$ 
production and asymmetry by the WA82 Collaboration~\cite{wa82}. Data were obtained using 
a $\pi^-$ beam with  W/Si and  W/Cu targets.

Once again, as evidenced in the 
figures, the result of the fit describes well the experimental data on both, production and 
production asymmetry. It has to be noted however that 
in fits to the $D^+$ WA82 data, a small contribution from the recombination 
of the hard QCD charm quark with the debris of the initial pion is necessary, opposite to what 
happens with the WA92 data. This behavior is due to the high value of the last experimental 
point, not seen in the WA92 $D^+$ data (See Fig.~\ref{fig96}). We also shown in the figure 
the asymmetry obtained from fits to WA92~\cite{wa92} and E769~\cite{alves1} data. As can be 
seen, both curves are similar and describe well the WA82 data on
production asymmetry. Note also that a null contribution from
recombination to $D^+$ production results in an asymmetry growing fast
than that shown by the full line in Fig.~\ref{fig97}, giving a better
agreement among data and model.

\begin{figure}[t]
\begin{minipage}{2.0cm}
\psfig{figure=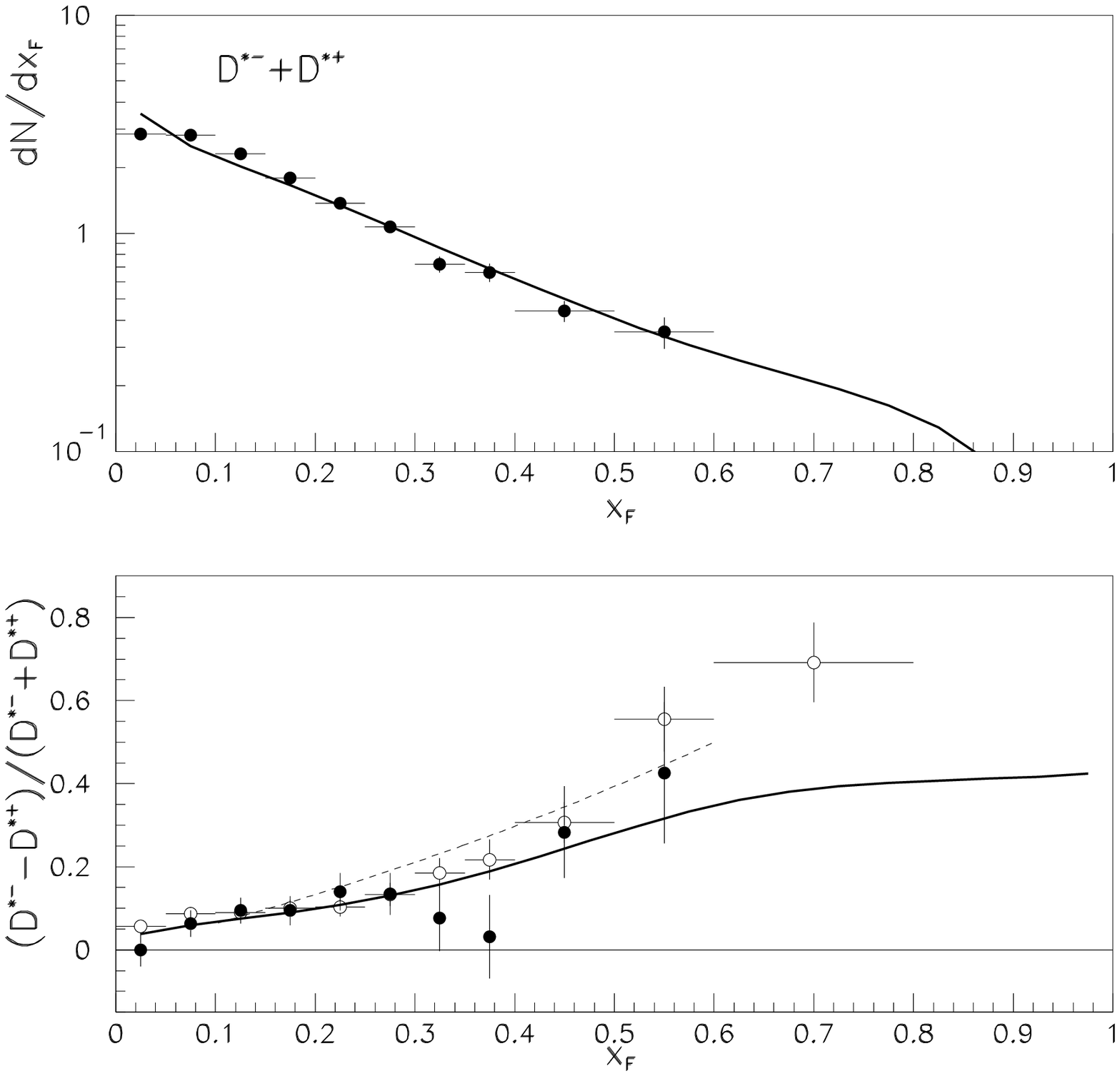,height=3.1in}
\end{minipage}
\begin{minipage}{7.5cm}
\psfig{figure=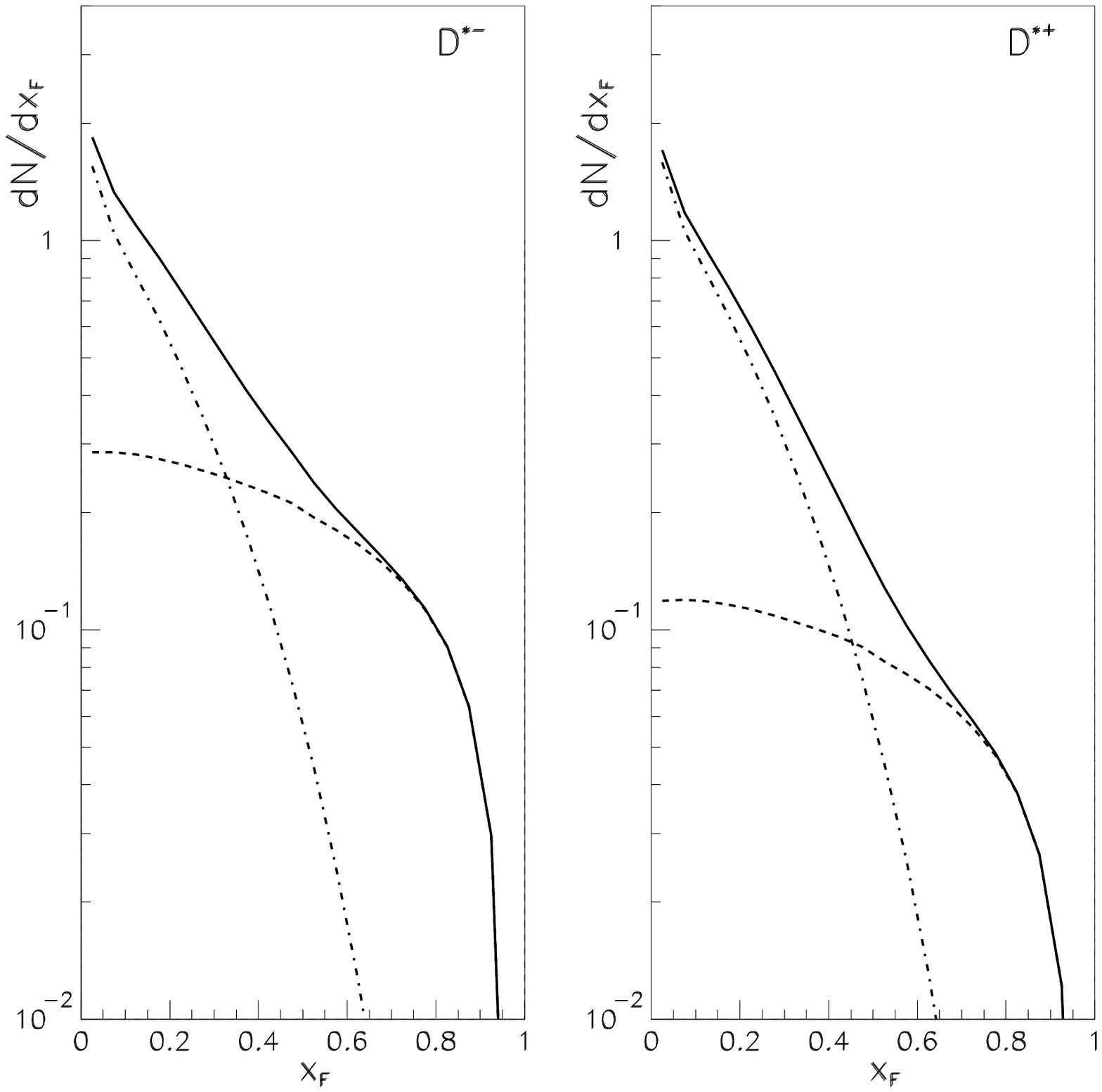,height=3.1in}
\end{minipage}
\caption{Left: $D^{*-}+D^{*+}$ production in $\pi^-~Nucleus$ interactions at 
500 GeV/$c$. Data from the E791 Collaboration~\cite{e791}. Full line: our model as in  
eq.~(\ref{comp1}). bottom:  $D^{*-}/D^{*+}$ production asymmetry. Full
line: our fit. Data from the E791 Collaboration ($D^*$: full
circles~\cite{e791}, $D^\pm$: open circles~\cite{e791-2}). 
Right: $D^{*-}$ and $D^{*+}$ $x_F$ distributions. Full line: $dN/dx_F$
as given by eq.~(\ref{comp1}). Dashed line: contribution from
recombination, point-dashed line: fragmentation.}
\label{fig98}
\end{figure}

In Fig.~\ref{fig98} we show the data and fits to the $D^*$(2010) production in 500 GeV/$c$ 
$\pi^- Nucleus$ interactions by the E791 Collaboration~\cite{e791}. The E791 measured 
the $D^{*-}+D^{*+}$ particle distribution as a function of $x_F$ as well as the $D^{*\pm}$ 
production asymmetry. In order to fit the data with our model of eq.~(\ref{comp1}), we fitted 
simultaneously the $D^{*-}+D^{*+}$ particle distribution and the production asymmetry. From the 
fits, we extracted the individual $D^{*-}$ and $D^{*+}$ particle distributions.
As can be seen in the figure, our curves agree well with the experimental data. Data on $D^\pm$ 
production asymmetry from the same experiment~\cite{e791-2} is superimposed on the $D^{*\pm}$ 
data. As shown in the figure, both charged $D$ and $D^*$ asymmetries
are the same within 
errors. The E769 Collaboration has also measured the $x_F$ $D^{*\pm}$ and $D^\pm$ distributions 
and asymmetry in 250 GeV/$c$ $\pi$-nucleon interactions~\cite{alves}. They reported that the 
behavior of the $x_F$ particle distributions are of the form $(1-x)^{2.9\pm0.4}$ and 
$(1-x)^{4.1\pm0.5}$ for leading and no-leading $D^*$ mesons
respectively in the region $0.1<x_F<0.6$. 
From these fits we extracted the asymmetry shown by the dashed line in Fig.~\ref{fig98}, which 
reproduces the same behavior of our fit to the $D^{*\pm}$ asymmetry measured by the 
E791 Collaboration. This gives support to the idea that the production mechanisms of the 
pseudoscalar $D^\pm$ and vector $D^{*\pm}$ mesons are the same, depending only on the 
flavor content of the initial and final particles. 

A comparison of all the data and curves on $D^\pm$ production asymmetries shows that the 
asymmetry seems to be independent of the collision energy.

\subsection{$D^0$ and $\overline{D}^0$ production}

\begin{figure}[t]
\begin{minipage}{2.0cm}
\psfig{figure=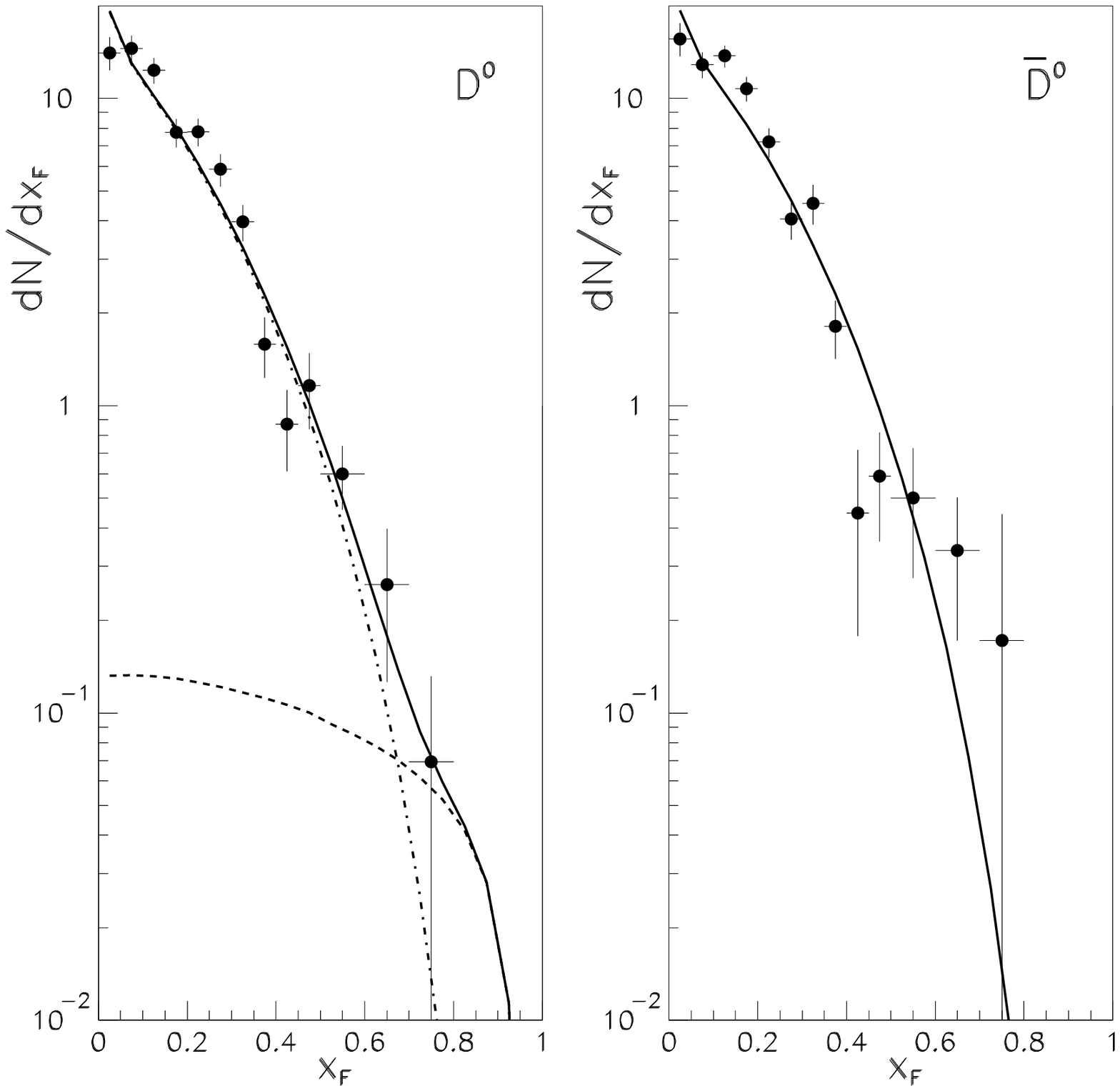,height=3.1in}
\end{minipage}
\begin{minipage}{7.5cm}
\psfig{figure=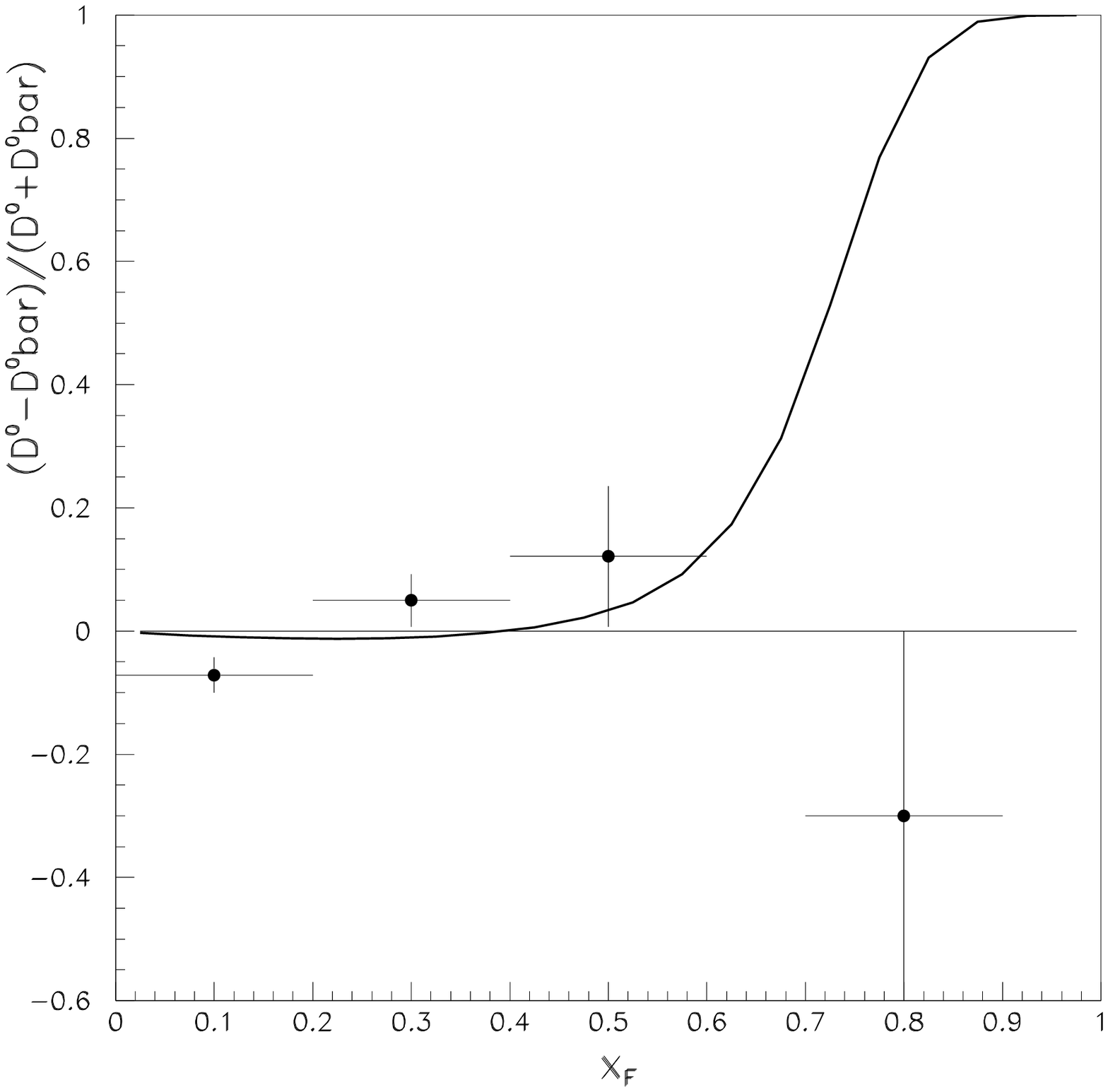,height=3.1in}
\end{minipage}
\caption{$D^0$ and $\overline{D}^0$ production by the WA92 Collaboration~\cite{wa92}. 
Left and middle panels: (full line) our model as in eq.~(\ref{comp1}) compared to experimental 
data. Dashed and point line: contributions from fragmentation and recombination respectively 
(See the text). Right: Production asymmetry. Model (full line) {\it vs} experimental data.}
\label{fig96b}
\end{figure}

In Fig.~\ref{fig96b} we display data on $D^0$ and $\overline{D}^0$ production by the 
WA92~\cite{wa92} experiment and compare to the model of eq.~(\ref{comp1}). As evidenced in the 
figure, the model describes quite well the experimental data. Notice that in $\pi^--Nucleus$ 
interactions, the $D^0~(c\bar u)$ is leading, then a positive asymmetry is expected. However, 
$\bar u$ valence quarks of the $\pi^-$ can anihilate easily with $u$ valence quarks 
in the Nucleus, then becoming unavailable to recombine with the charm quark to produce a $D^0$ 
state. Consequently both $D^0$ and $\overline{D}^0$ are mainly produced by charm quark 
fragmentation at the same rate and the asymmetry is largely suppressed as noted in 
Refs.~\cite{magnin,bedniakov}. Nevertheless, at large $x_F$ there is an excess of both $D^0$ 
and $\overline{D}^0$ which still requires a small contribution from recombination.  

Table~\ref{table1} lists the results of the fits to experimental data
on $D$ production and asymmetries.
\begin{table}[h]
\begin{center}
\begin{tabular}{|c|c|c|c|c|}
\hline\hline
Experiment & Particle & a & b & $\chi^2$/d.o.f. \\
\hline
WA82 & $D^-$ & 14269.27 $\pm$ 1116.82 & 918$\pm$ 140.57 & 3.570 \\
\hline
WA82 & $D^+$ & 18649.11 $\pm$ 1828.77 & 311.99 $\pm$ 128.22 & 2.368 \\
\hline
WA92 & $D^-$ & 292.64 $\pm$ 14.59 & 9.59 $\pm$ 2.43 & 3.075 \\
\hline
WA92 & $D^+$ & 302.46 $\pm$ 13.92 & 0. $\pm$ 1.34 & 1.698 \\
\hline
E791 & $D^{*-}$ & 41.54 $\pm$ 2.14 & 4.86 $\pm$ 0.54 &  4.613 \\
\hline
E791 & $D^{*+}$ & 48.40 $\pm$ 2.17 &  1.97 $\pm$ 0.44 & 4.613 \\
\hline
WA92 & $D^0$ & 630.74 $\pm$  29.32 & 4.17 $\pm$ 3.26 & 3.137 \\
\hline
WA92 & $\bar{D}^0$ & 733.51 $\pm$ 37.25 & 0. $\pm$ 3.90 & 4.370 \\
\hline\hline
\end{tabular}
\caption{Coefficients obtained in fits to experimental data on $D$
  meson production. The $\chi^2$/d.o.f. of the E791 data on
  $D^*$ production is for a simultaneous fit to the $D^{*-}+D^{*+}$
  $x_F$ distribution and asymmetry.}
\label{table1}
\end{center}
\end{table}
As shown in Table~\ref{table1}, the $\chi^2$ of the fits are in general
larger than desirable. This somewhat unpleasant aspect of the fits is
mainly due to two reasons, 1) the bad quality of experimental data
exhibiting large error bars in most cases and, 2) the small number of parameters of
the model, which do not allow to describe the details of the data but
the more or less gross features of them. 

\subsection{$\Lambda_c^\pm$ production}

\begin{figure}[t]
\begin{minipage}{2.0cm}
\psfig{figure=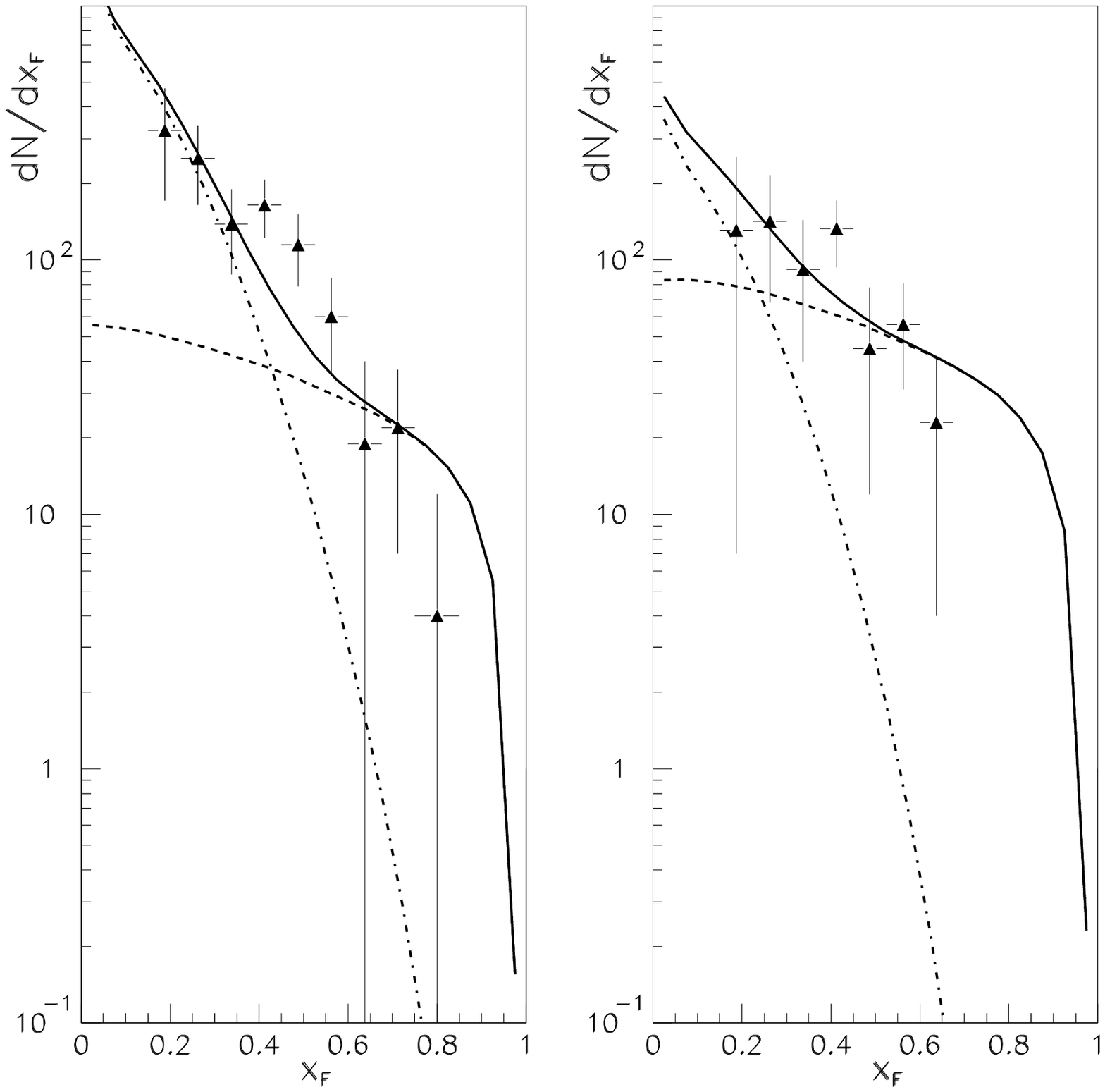,height=3.1in}
\end{minipage}
\begin{minipage}{7.5cm}
\psfig{figure=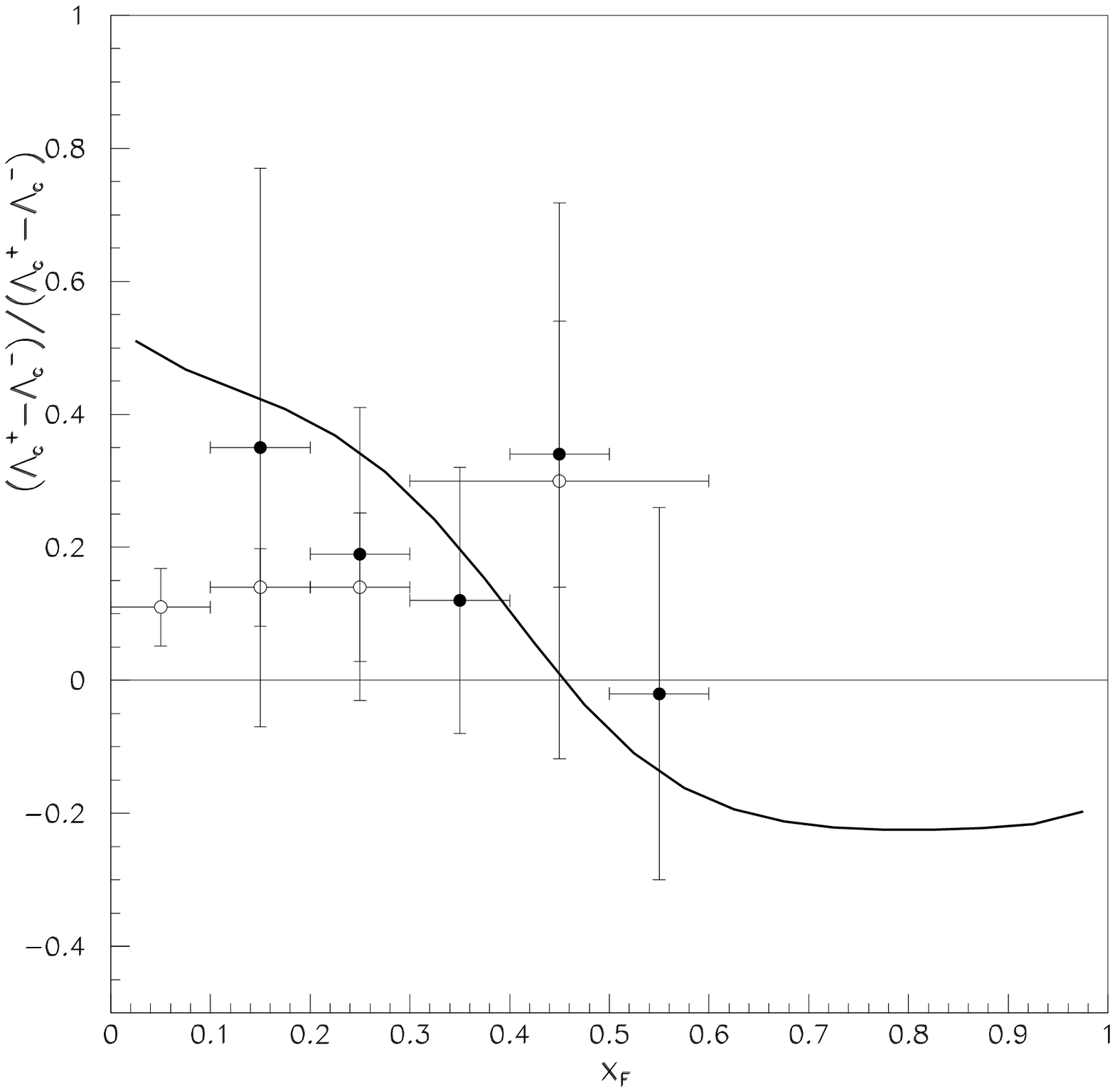,height=3.1in}
\end{minipage}
\caption{$\Lambda_c^+$ (left) and $\Lambda_c^-$ (middle) production in $\pi^-~Nucleus$ 
interactions at 600 GeV/$c$. Data from the SELEX Collaboration~\cite{selex}. 
Full line: our model as in  eq.~(\ref{comp1}). Right: $\Lambda_c^+/\Lambda_c^-$ production 
asymmetry. Data from Ref.~\cite{selex} (full circles) and
Ref.~\cite{e791-baryon} (open circles). Full line: result of our fit.}
\label{fig99}
\end{figure}

In Fig.~\ref{fig99}, we display the data and fit results for $\Lambda_c$ production 
and production asymmetry obtained in 600 GeV/$c$ $\pi^- Nucleus$ interactions by the 
SELEX Collaboration~\cite{selex}. We display also the E791~\cite{e791-baryon} data on $\Lambda_c$ 
production asymmetry in $\pi^--Nucleus$ interactions at 500 GeV/$c$.

Here we fitted the $\Lambda_c^+$ and $\Lambda_c^-$ particle 
distributions and obtained the asymmetry from our fitting functions. 
As can be seen in the figures, 
our theoretical curves agree well with the experimental data. 

\begin{figure}[t]
\centerline{\psfig{figure=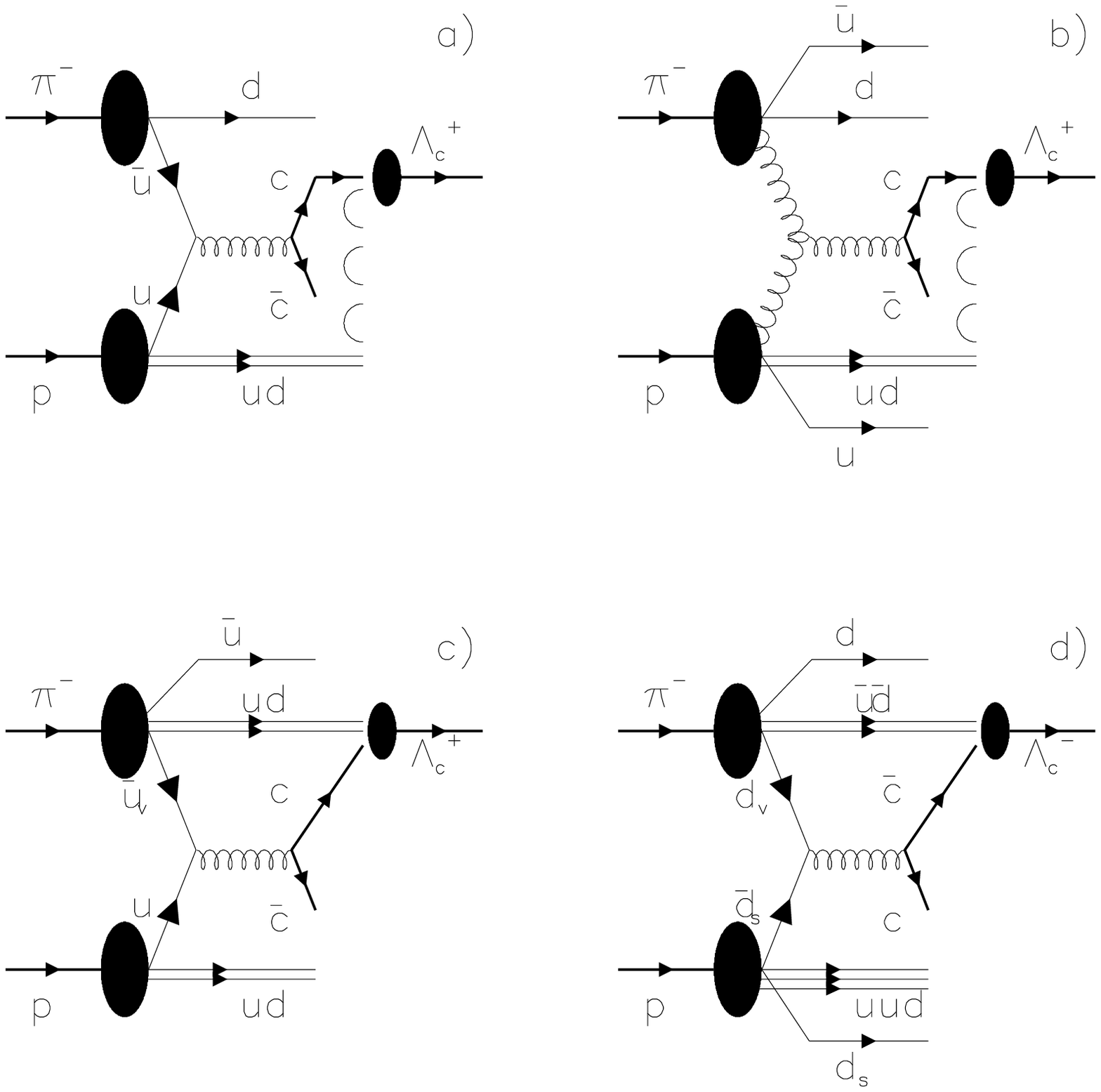,height=3.5in}}
\caption{Some typical diagrams for $\Lambda_c^+$ and $\Lambda_c^-$ production in 
$\pi^-~Nucleus$. a) and b) $\Lambda_c^+$ 
production by color string fragmentation. The $c$ quark forms a baryon-like string with 
a diquark in the proton. The $\bar c$ quark forms a meson-like string with the remaining valence 
quark of the $\pi^-$. c) and d) $\Lambda_c^\pm$ production by recombination.}
\label{fig101}
\end{figure}

It is interesting to note that, although both $\Lambda_c^+~(udc)$ and 
$\Lambda_c^-~(\bar{u}\bar{d}\bar{c})$ are leading in 
$\pi^-~(\bar{u}d)$-Nucleon interactions, there is a significant difference in their 
$x_F$ distributions. 
In fact, as noted in Ref.~\cite{anjos}, valence $\bar u$ quarks in the $\pi^-$ can annihilate 
easily with (valence) $u$ quarks in nucleons but, valence $d$ quarks of the pion cannot, thus 
favoring the $\Lambda_c^+$ production over the $\Lambda_c^-$ one. This is the case of diagrams 
c) and d) in Fig.~\ref{fig101}. However, not only the recombination mechanism contributes to 
the $\Lambda_c^\pm$ production asymmetry in $\pi^-~Nucleon$ interactions. Color 
string fragmentation also gives a contribution. Looking at the diagrams a) and b) in 
Fig.~\ref{fig101} we note that also in this case $\Lambda_c^-$ production is disfavored. 
An analogue diagram to that of Fig.~\ref{fig101} a), with the valence $d$ quark in the $\pi^-$ 
annihilating with a $\bar d$ quark in the proton and a $\bar c$ quark produced in the forward 
($x_F > 0$) region giving rise to a $\Lambda_c^-$ is suppressed by at least two reasons:
1) $\bar u_\pi ~-~u_p$ anihilation is favored over $d_\pi~-~\bar d_p$ because in the first case 
both are valence quarks and 2) a $\bar c$ quark produced in the forward region has to form a 
baryonic color string with a $\bar u \bar d$ sea diquark in the proton in order to produce a 
$\Lambda_c^-$. The same situation is present in the case of Fig.~\ref{fig101} d).

\begin{figure}[t]
\begin{minipage}{2.0cm}
\psfig{figure=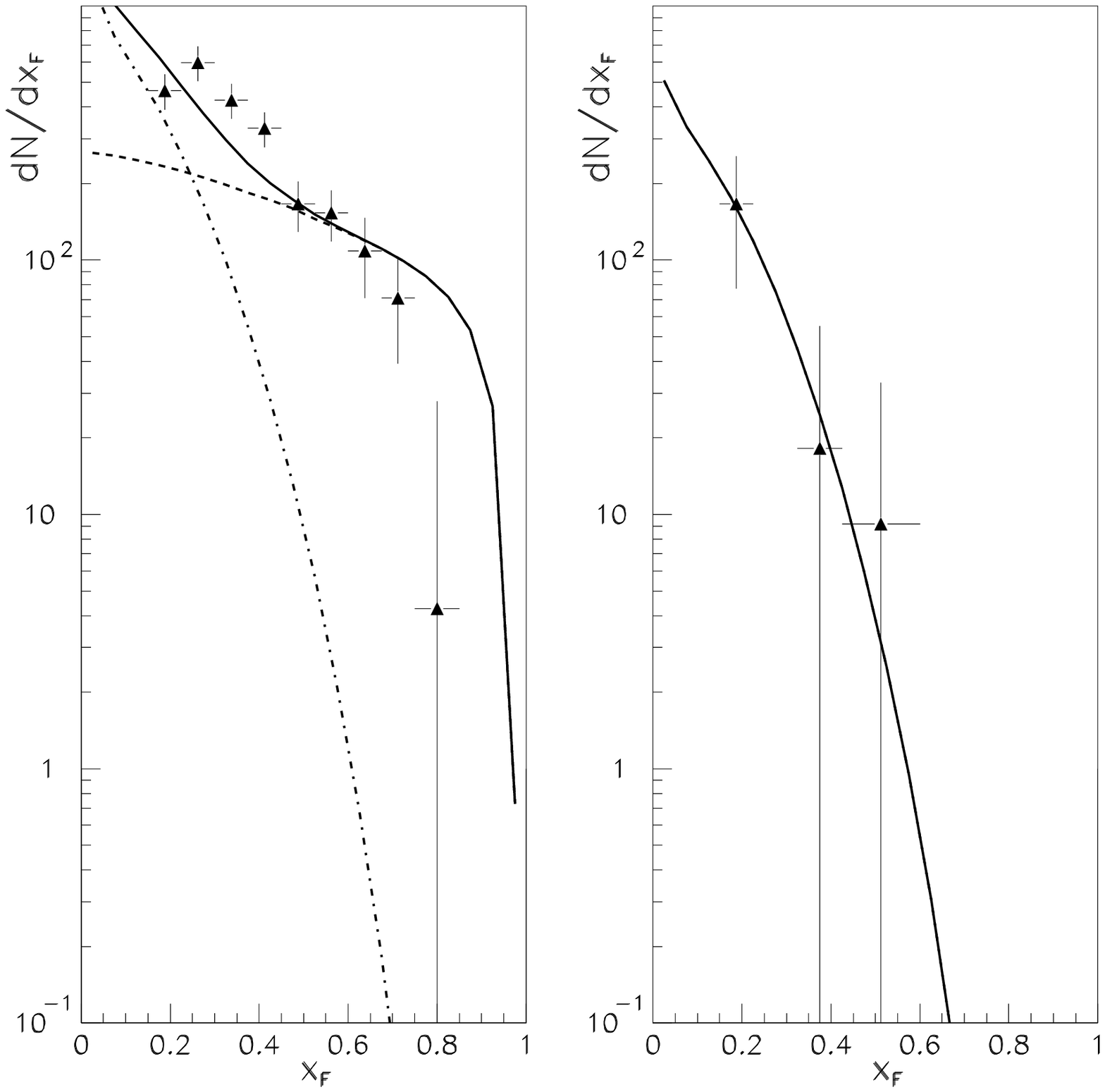,height=3.1in}
\end{minipage}
\begin{minipage}{7.5cm}
\psfig{figure=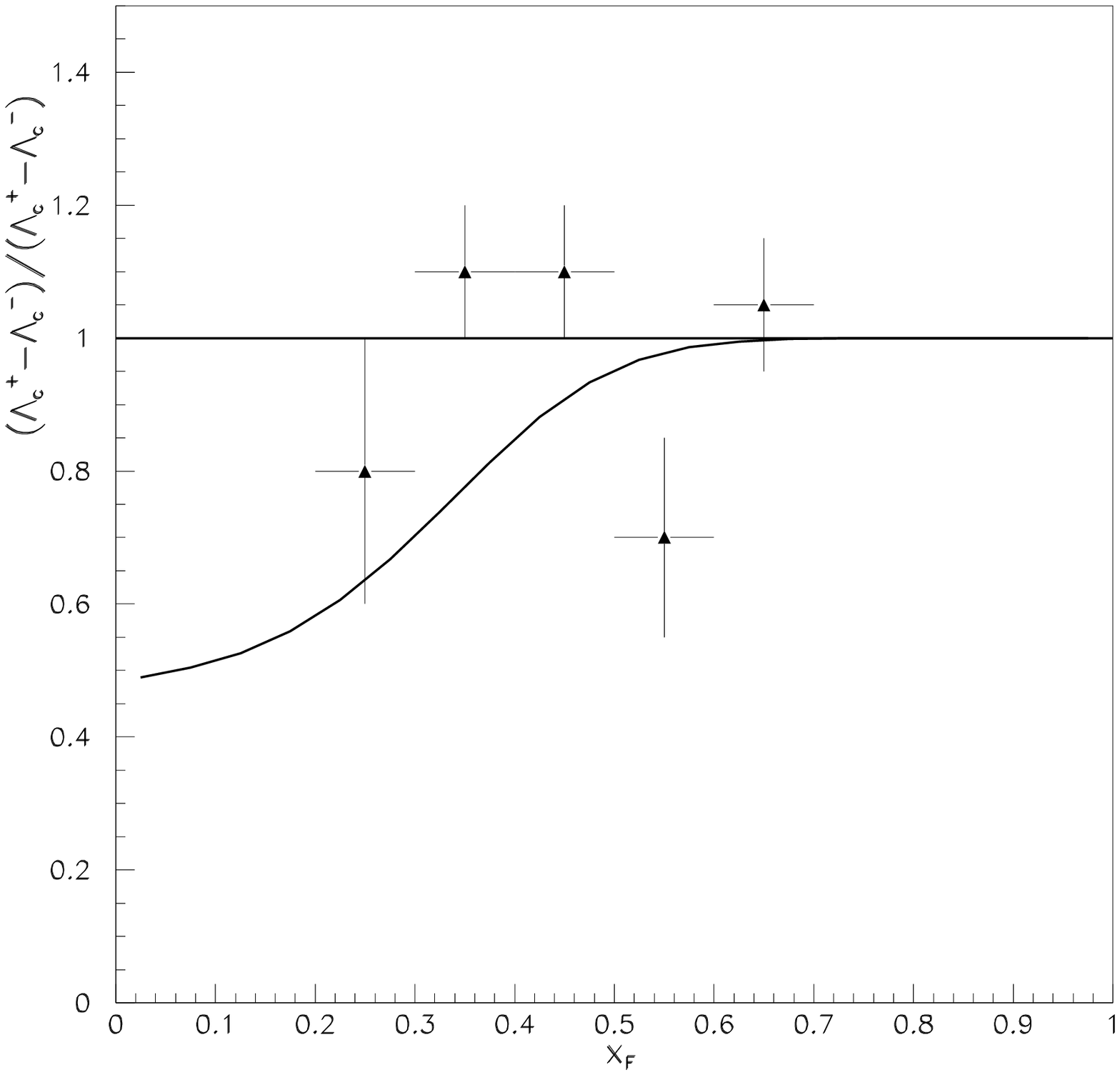,height=3.1in}
\end{minipage}
\caption{$\Lambda_c^+$ (left) and $\Lambda_c^-$ (middle) production in $p~Nucleus$ 
interactions at 600 GeV/$c$ beam energy. Data from the SELEX Collaboration~\cite{selex}. 
Full line: our model as in  eq.~(\ref{comp1}). Right: $\Lambda_c^+/\Lambda_c^-$ production 
asymmetry. Data from Ref.~\cite{selex}. Full line: result of our fit.}
\label{fig100}
\end{figure}

Finally, in Fig.~\ref{fig100} we show the data and fits to $\Lambda_c^\pm$ production and 
production asymmetry in $p~N$ interactions at 600 GeV/$c$~\cite{selex}. In order to get the 
theoretical curves, we fitted both the $\Lambda_c^+$ and $\Lambda_c^-$ distributions 
as a function of $x_F$ and calculated the asymmetry from the curves obtained in fits. 

\begin{figure}[t]
\centerline{\psfig{figure=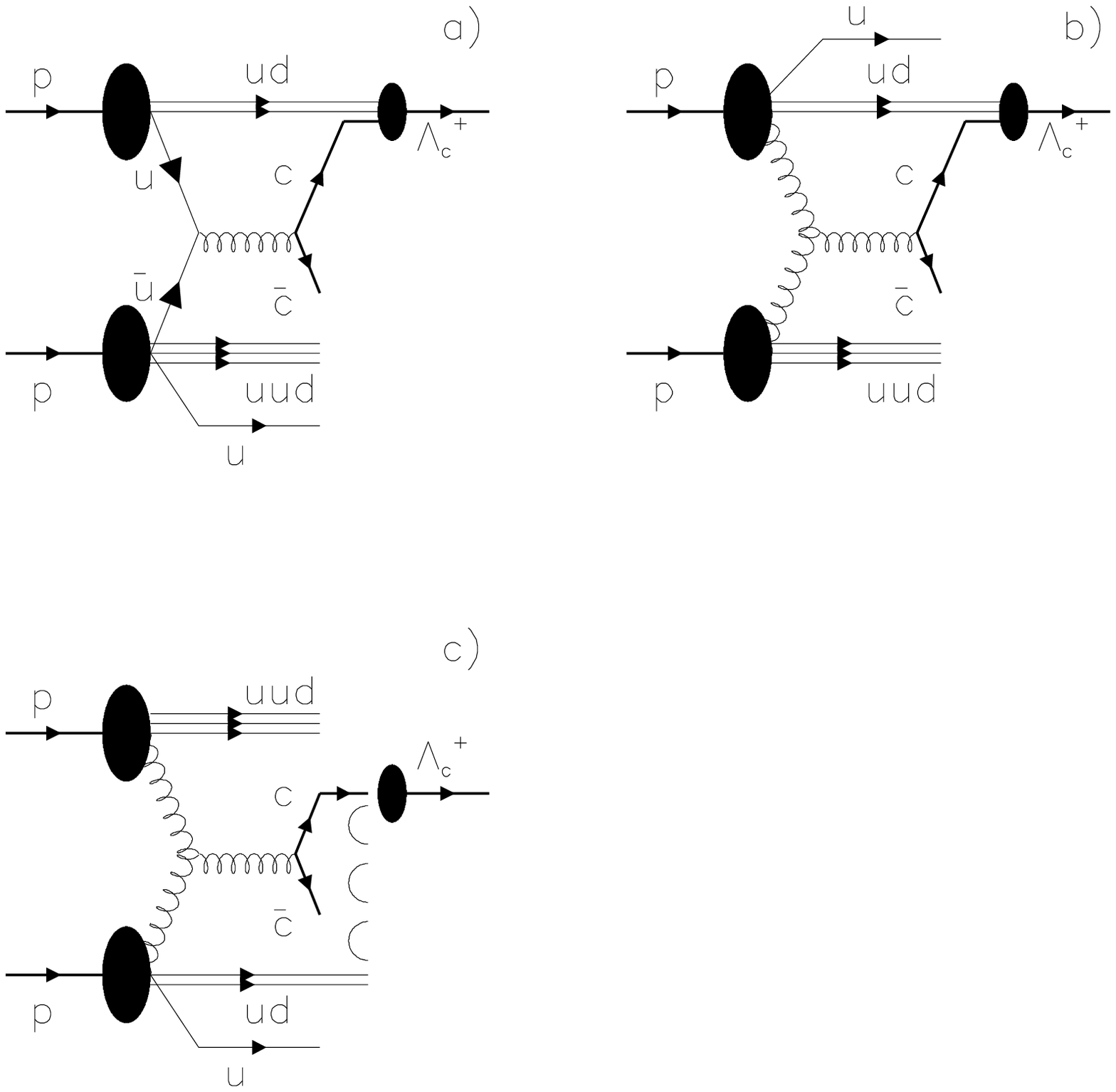,height=3.1in}}
\caption{Typical diagrams for $\Lambda_c^+$ and $\Lambda_c^-$ production in $p~Nucleus$. 
a) and b) $\Lambda_c^+$ 
production by recombination. c) $\Lambda_c^+$ production by color string fragmentation.}
\label{fig102}
\end{figure}

For the $\Lambda_c^-$ we only use the first term in eq.~(\ref{comp1}), 
neglecting any contribution from recombination. The reason is that in order to get a 
$\Lambda_c^-$ baryon from recombination, the perturbatively produced $\bar c$ quark has to 
recombine with a $\bar u \bar d$ diquark formed from sea quarks in the initial proton, which 
is less favorable than the recombination of a $c$ quark with a $ud$ valence diquark (see 
Fig.~\ref{fig102}).

As in the case of $\Lambda_c^\pm$ production in $\pi^-~-~Nucleon$ interactions, $\Lambda_c^+$ 
production is favored over the $\Lambda_c^-$, having not only a different 
shape in their $x_F$ distributions but also a different global normalization, as shown in 
Fig.~\ref{fig102}. This is a distinctive feature of charm baryon production not seen in 
charm meson production, where both, particle and antiparticle $x_F$ distributions begin at 
more or less the same point at $x_F \sim 0$.

Concerning the data on the production asymmetry in $p~Nucleus$
interactions by SELEX, it is hard to understand, however, how they
obtain three points with asymmetry bigger than one. Note that 
from the definition of the asymmetry follows that it is bounded to be between $1$
and $-1$.

Table~\ref{table2} shows the results of fits to experimental data on
$\Lambda_c^\pm$ production. Once again, as in the case of meson
production, $\chi^2$ of fits are far away from their desirable
values. However, note that in the case of baryon production,
experimental data exhibits larger error bars than in the case of mesons. 

\begin{table}[h]
\begin{center}
\begin{tabular}{|c|c|c|c|c|}
\hline\hline
Experiment & Particle & a & b & $\chi^2$/d.o.f. \\
\hline
$\pi^-~-N$ & $\Lambda_c^+$ & 76511.84 $\pm$ 18661.76 & 990.78 $\pm$
303.93 & 1.602 \\
\hline
$\pi^-~-N$ & $\Lambda_c^-$ & 22625.39 $\pm$ 18946.02 & 1682.04 $\pm$
478.72 & 0.786 \\
\hline
$p~-N$ & $\Lambda_c^+$ & 77904.03 $\pm$ 16361.06 & 5226.22 $\pm$
608.73 & 4.077 \\
\hline
$p~-N$ & $\Lambda_c^+$ & 32204.85 $\pm$ 19726.59 & 0.00 $\pm$ 688.32 &
0.093 \\
\hline\hline
\end{tabular}
\caption{Coefficients obtained from fits to data on $\Lambda_c^\pm$
  production from the SELEX Collaboration.}
\label{table2}
\end{center}
\end{table}

\section{Conclusions}

We have shown that using a simple model based on perturbative QCD for charm quark production 
and on well known hadronization mechanisms, namely fragmentation and recombination, the available 
experimental data on charm hadron production can be well
described. The idea of describing charm hadron 
production by means of fragmentation and recombination of perturbatively produced charm quarks 
is not new. It has been discussed by the first time in Ref.~\cite{kartvelishvili}, but never has 
been extensively tested against most of the available experimental data. 

Our analysis shows that intrinsic charm is unnecessary to describe consistently 
the experimental data on both charm particle $x_F$ distributions and charm particle production 
asymmetries. This was evident once SELEX data on $\Lambda_c$ baryons and antibaryons became 
available, as noted in Ref.~\cite{anjos}.

Parameters in the model of eq.~(\ref{comp1}) represent the unkowns associated to 
the relative fractions of the fragmentation and recombination contributions, but also include the 
uncertainties coming from the non-perturbative contributions to the hadronization process. 
It is conceivable that with more abundant and precise experimental data, the behavior of this 
parameters with respect to the reaction and reaction energy can be fixed, allowing in this way 
a more predictive power of the model.

In Refs.~\cite{magnin,anjos} another model including the recombination of charm quarks 
with the debris of the initial particles was considered. However, in this model, recombining 
charm quarks were considered as part of the structure of the initial particles. The fact that 
both, the model discused here and the model of Refs.~\cite{magnin,anjos}, are able to describe 
well the experimental data is due to the fact that the momentum distribution of perturbatively 
produced charm quarks is similar to the sea charm quark distribution in hadrons. It is easy to 
see that this is a consequence of the fact that in both cases the origin of the charm quarks is 
gluon splitting, then its momentum distributions must be similar.

In addition, we have shown that factorization is broken as long as the structure of the initial 
colliding particles has to be taken into account in order to describe the hadronization of charm 
quarks. However, it does not mean that new unknowns are added to the problem. There still 
exist a consistent way to calculate hadronization within the framework of the recombination 
model.

Finally, we would like to stress that not only recombination, but also fragmentation, 
contributes to the observed baryon production asymetries. In recombination, the asymmetry is 
a consequence of the sharing of partons among the initial and final particles. In fragmentation, 
the asymmetry is due to the parton content of the initial particles. Note that, as long as 
initial particles are baryons and mesons, it is easier to have mesonic and baryonic color string, 
but no ``anti-baryonic'' ones, thus baryon production is favored over antibaryon production. 
This effect does not appear in meson-antimeson production since they are formed mainly from 
mesonic strings. If antibaryons were used as initial particles, fragmentation would favor the 
production of antibaryons, producing the opposite effect. It is interesting to note also that 
recombination gives a small contribution only noticeable at large $x_F$ (of the order 
of $x_F\sim 0.6-0.7$) to the $D$ meson $x_F$ distributions while for baryons its contribution 
becomes important for $x_F\sim 0.4$.

\section*{Acknowledgments}

J. Magnin would like to thanks the warm hospitality at the Physics Department, Universidad de 
los Andes, where part of this work was done. L.M. Mendoza-Navas would like to thanks the warm 
hospitality at CBPF during his stay for the completion of the work. This works was supported by 
Fundaci\'on para la Promoci\'on de la Investigaci\'on y la Tecnolog\'{\i}a del Banco de la 
Rep\'ublica (Colombia) under contract Project No.: 1376; CNPq, the Brazilian Council for 
Science and Technology and FAPERJ (Brazil) under contract Project No.:
E-26/170.158/2005.

\end{document}